\documentclass[preprintnumbers]{revtex4}
\UseRawInputEncoding
\usepackage{amssymb}
\usepackage{amsmath}
\usepackage{graphicx}
\usepackage{dcolumn}
\usepackage{bm}
\usepackage{subfigure}
\usepackage{color}

\begin{document}

\title{Barrow's non-linear charged Anti-de Sitter black hole and stability}

\author{Yun-Zhi Du$^{1,2}$\footnote{the corresponding author} and Qiang Gu$^{1}$}
\affiliation{Shanxi Datong University, Datong 037009, China\\
State Key Laboratory of Quantum Optics and Quantum Optics Devices, Shanxi University, Taiyuan, Shanxi 030006, China}

\thanks{\emph{e-mail:duyzh13@lzu.edu.cn, gudujianghu23@163.com}}

\begin{abstract}
As we know that the horizon area of a black hole will increase when it absorbs matters. While based on Barrow's idea of fractal black hole horizon, ones [Phys. Lett. B 831 (137181) 2022] had proposed that for a spherically fractal structure the minimal increase of the horizon area is the area of the smallest bubble sphere. And the corresponding black hole entropy is of a logarithmic form, which is similar to that of Boltzmann entropy under a certain condition. Based on these, we re-derive the entropy of the Barrow's Einstein-power-Yang-Mills (EPYM) AdS black hole, and calculate the temperature and heat capacity of the Barrow's EPYM AdS black hole. There exists an interesting phenomena that the ratio between the Barrow's temperature and the Hawking temperature of the EPYM AdS black hole is fully consistent with that of other Schwarzschild-like black holes. The Barrow's temperature and Hawking temperature with the certain range of $\Lambda$ are monotonically increasing and the corresponding heat capacities are all positive, which means these black holes are thermodynamically stable. Besides, for the Barrow's EPYM AdS black hole its heat capacity has a Schottky anomaly-like behavior, which may reflect the existence of the discrete energy level and the microscopical degree of freedom.

\end{abstract}

\maketitle

\section{Introduction}
As we known that there arises the non-renormalizability which is a well-known puzzle when gravity is combined with quantum theory. In $1916$ Einstein published the general relativity theory in which a concept about acceleration differences between non-inertial frames was proposed. It sheds light on the idea that the space and the time are both distinct features, and the spacetime is curved. However, the spacetime curvature depends on the presence of gravity, which is also capable of wrapping light, radio waves and many other things. In $1939$, Oppenheimer and Snyder published a paper titled as ``On Continuing Gravitational Contraction'' on black holes \cite{Oppenheimer1939}. They predicted the continued contraction of a star, forming a body with an extreme attraction force that could not escape even light from it. However, Einstein established the general relativity and predicted the presence of gravitational waves, although he did not have any experimental evidence to prove the existence of gravitational waves. Over nearly half of a century, when LIGO physically detected the undulations in spacetime induced by gravitational waves, the so-called ``GW150914'' phenomenon, is a proof of Einstein's theory \cite{Abbatial2016}. This is the first clear observation of a pair of black holes that merge to form a single black hole (BH). Black holes are possibly the most ideal thermal structures that have played an increasingly important role in the general relativity interdisciplinary fields, the information theory, the quantum mechanics, and the statistical physics.

The four laws of BH mechanics formalized its behavior and the microscopic properties, closely implies the thermal properties of BHs \cite{Bardeen1973,Wald2001}. After the work of Bekenstein, Stephen Hawking was the first to demonstrate the presence of BH radiations, which is defined as a tunneling process caused by vacuum fluctuations near the horizon. A small amount of heat called Hawking Radiation recorded by Hawking \cite{Hawking1975} is radiated due to irregularities of quantum mechanics. BHs were found to possess thermodynamic properties such as entropy and temperature in the Pioneering work \cite{Bekenstein1972,Bekenstein1973}. These findings show that a BH is a thermodynamic system with a Hawking temperature proportional to its surface gravity on the horizon area and obeys the four BH thermodynamic laws. AdS BH thermodynamics is the most impressive instrument for the study of quantum gravity in spacetime containing horizon \cite{Birrell1982}. The quantum scenario for Hawking radiation provides a valuable insight that a black hole temperature is proportional to its horizon area and that its entropy is related to its area gravity \cite{Abreu2020}. This analysis revealed that there is a strong relationship between thermodynamics and gravity. The Barrow entropy \cite{Barrow2020} arises from the notion that the surface of a BH can be altered as a result of quantum gravitational effects. Recently Barrow has shown that the area of a BH can be represented as a discontinuous and possibly fractal structure of the geometry of the horizon \cite{Barrow2020}. Starting from the idea of ``Koch snowflake'' Barrow proposed a model in which the event horizon of a Schwarzschild black hole radius is surrounded by $N$ smaller spheres whose radii have a ratio $\lambda$ with respect to the original sphere. After $n$ iterations, the radius is $r_{n+1}=\lambda r_n$. By adding smaller spheres to the surface, the total area and volume after an infinite number of steps should be the sum of all the intricate structures, which can be finite and infinite. The corresponding entropy can be very large. Subsequently, the Barrow's idea has been further extended to the investigation of the dark energy \cite{Saridakis2020,Moradpour2020}, the cosmology \cite{Salehi2023,Komatsu2024,Okcu2024} and black hole thermodynamics \cite{Abreu2020a,Ma2022,Ladghami2024,Rani2023}.

In Ref. \cite{Ma2022} the authors proposed a new way to investigate the effect of a minimal length on the entropy and the stability of black holes with fractal structures by considering the generalized uncertainty principle. Based on this idea, we apply the corresponding Barrow's entropy formula to the Einstein-power-Yang-Mills (EPYM) AdS black hole, and analyze the stability of Barrow's black holes. In Sec. \ref{scheme2}, we briefly review the thermodynamic quantities and stability of the standard EPYM AdS black hole. In Sec. \ref{scheme3}, we review Barrow's fractal black hole and the Barrow's entropy formula with the consideration of the generalized uncertainty principle. Then based on the fractal structure we calculate the temperature and heat capacity of the AdS black hole with the non-linear source. A brief summary is given in Sec. \ref{scheme4}.

\section{Non-linear charged AdS black hole Solution}
\label{scheme2}

\begin{figure}[htp]
\includegraphics[width=0.45\textwidth]{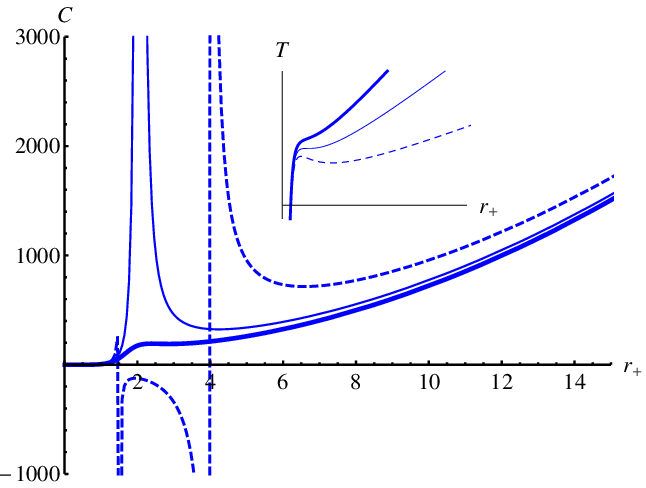}
\caption{Plots of $T,C-r_+$ with the parameters $q=0.85$, $\gamma=0.8$, and $\Lambda=-0.04474$ (the thicked solid lines), $\Lambda=-0.08947$ (the thin solid lines), $\Lambda=-0.1342$ (the dashed lines).}\label{C-T-r}
\end{figure}

The action for four-dimensional Einstein-power-Yang-Mills (EPYM) gravity with a cosmological constant $\Lambda$ was given by \cite{Zhang2015,Corda2011,Mazharimousavi2009,Lorenci2002} ($8\pi G=1$)
\begin{eqnarray}
I=\frac{1}{2}\int d^4x\sqrt{g}
\left(R-2\Lambda-\mathcal{F}^\gamma\right)
\end{eqnarray}
with the Yang-Mills (YM) invariant $\mathcal{F}$ and the YM field $F_{\mu \nu}^{(a)}$
\begin{eqnarray}
\mathcal{F}&=&\operatorname{Tr}(F^{(a)}_{{\mu\nu}}F^{{(a)\mu\nu}}),\\
F_{\mu \nu}^{(a)}&=&\partial_{\mu} A_{\nu}^{(a)}-\partial_{\nu} A_{\mu}^{(a)}+\frac{1}{2 \xi} C_{(b)(c)}^{(a)} A_{\mu}^{(b)} A_{\nu}^{(c)}.
\end{eqnarray}
Here, $\operatorname{Tr}(F^{(a)}_{\mu\nu}F^{(a)\mu\nu})
=\sum^3_{a=1}F^{(a)}_{\mu\nu}F^{(a)\mu\nu}$, $R$ and $\gamma$ are the scalar curvature and a positive real parameter, respectively; $C_{(b)(c)}^{(a)}$ represents the structure constants of three-parameter Lie group $G$; $\xi$ is the coupling constant; and $A_{\mu}^{(a)}$ represents the $SO(3)$ gauge group Yang-Mills (YM) potentials defining by the Wu-Yang (WY) ansatz \cite{Balakin2016,Balakin2007,Balakin2007a}. For this system, the EPYM black hole solution with the negative cosmological constant $\Lambda$ in the four-dimensional spacetime is obtained by adopting the following metric \cite{Yerra2019}:
\begin{eqnarray}
d s^{2}=-f(r) d t^{2}+f^{-1} d r^{2}+r^{2} d \Omega_{2}^{2},\\
f(r)=1-\frac{2 M}{r}-\frac{\Lambda}{3} r^{2}+\frac{\left(2 q^{2}\right)^{\gamma}}{2(4 \gamma-3) r^{4 \gamma-2}},
\end{eqnarray}
where $d\Omega_{2}^{2}$ is the metric on unit $2$-sphere with volume $4\pi$ and $q$ is the YM charge. Note that this solution is valid for the condition of the non-linear YM charge parameter $\gamma\neq0.75$, and the power YM term holds the weak energy condition (WEC) for $\gamma>0$ \cite{Corda2011}. In the extended phase space, $\Lambda$ was interpreted as the thermodynamic pressure $P=-\frac{\Lambda}{8\pi}$. For $\gamma=1$, the YM charge term in the metric has the same form as the Maxwell charge term for the Einstein-Maxwell-Yang-Mills (EMYM) theory. Thus the contribution of the YM charge term on the thermodynamic property should be same as that of the Maxwell charge term in EMYM theory. The only difference is that they have different gauge groups: the YM field is of $SO(3)$, while the EM field is of $U(1)$. These comments are consistent with that in Ref. \cite{Moumni2018}. The black hole event horizon locates at $f(r_+)=0$. The parameter $M$ represents the ADM mass of the black hole and it reads
\begin{eqnarray}
M(r_+, q, \gamma)=\frac{r_+}{2}\left[1-\frac{\Lambda r_+^2}{3}+\frac{\left(2 q^{2}\right)^{\gamma}}{2(4 \gamma-3)r_+^{4\gamma-2}}\right].\label{M}
\end{eqnarray}
And in our set up it is associated with the enthalpy of the system. The black hole temperature and Hawking entropy were given by \cite{Zhang2015}
\begin{eqnarray}
T=\frac{1}{4 \pi r_{+}}\left(1-\Lambda r_{+}^{2}-\frac{\left(2 q^{2}\right)^{\gamma}}{2 r_{+}^{4 \gamma-2}}\right),~~~~~S=\pi r_{+}^{2}.    \label{T}
\end{eqnarray}
In order to probe the stability of the EPYM AdS black hole, we present the heat capacity at the constant charge $q$, which is a counterpart of the heat capacity at the constant volume, as the following form
\begin{eqnarray}
C=T\frac{\partial S}{\partial T}=-\frac{2\pi r_+^2\left(1-\Lambda r_+^2-\frac{\left(2 q^{2}\right)^{\gamma}}{2 r_{+}^{4 \gamma-2}}\right)}{1+\Lambda r_+^2-\frac{(2\gamma-1)\left(2 q^{2}\right)^{\gamma}}{2 r_{+}^{4 \gamma-2}}}.\label{C-stand}
\end{eqnarray}
The behaviours of the temperature and the heat capacity for the EPYM AdS black hole are shown in Fig. \ref{C-T-r}. The results indicate that there exist a critical value of the cosmological constant $\Lambda_c=-\frac{\gamma-1/2}{(2q^2\gamma^2)^{\frac{\gamma}{2\gamma-1}}
(4\gamma-1)^{\frac{1}{2\gamma-1}}}$, when $\Lambda\geq\Lambda_c$ the temperature is monotonically
increasing with $r_+$, the heat capacity that has a Schottky Anomaly-like behaviour and it is always positive. Below the critical value, the temperature have some extremes, and the corresponding heat capacity is positive only in the interval where the slope of the temperature is positive. This configuration corresponds to the first-order phase transition between the high-potential black hole and the low-potential one as shown in Ref. \cite{Du2021}. In the next part, we will investigate the temperature and heat capacity of the Barrow's EPYM AdS black hole.

\section{Barrow's EPYM AdS entropy and stability}
\label{scheme3}

From Barrow's idea, the fractal structure of Schwarzschild black hole horizon is that there are many smaller spheres that can attach on the black hole horizon, then more smaller spheres attaching to these spheres and so on. At each step there are $N$ spheres and the horizon radius is $\lambda$ times smaller than that of the sphere in the previous step. For no cut off at some small finite scale, the corresponding area of the horizon and volume of the black hole were given in Ref.
\begin{eqnarray}
A_{\infty}=4\pi r^2_h\sum^{\infty}_{n=0}\left(N\lambda^2\right)^n,~~
V_{\infty}=\frac{4\pi r^3_h}{3}\sum^{\infty}_{n=0}\left(N\lambda^3\right)^n,
\end{eqnarray}
where $r_h$ is the black hole horizon radius and $\lambda$ is the ratio of radii with respect to the original sphere. As $\lambda^{-2}<N<\lambda^{-3}$ the black hole horizon area will be infinite, while the volume is finite. On the other hand, Barrow proposed that the entropy can take the form as $S\approx (A_h/A_{pl})^{1+\triangle/2}$ with $0<\triangle<1$. Here $A_h$ stands for the standard horizon area and $A_{pl}$ is the Planck area. As $\triangle=0$ it reduces to the standard black hole entropy, and it corresponds to one of the most intricate horizon when $\triangle=1$.

In the frame of the generalized uncertainty principle, the lowest limit of the length scale is the Planck length $l_{p}$. The radius of the smallest sphere in the fractal structure can not exactly be $l_p$. From the recurrence relation $r_{n+1}=\lambda r_h$, at some cut-off step $n_1$ it only needs to satisfy $r_{n1}\geq l_p,~\lambda r_{n1}<l_p$. In this case, the Barrow's black hole entropy was given in Ref. \cite{Ma2022}
\begin{eqnarray}
S_B=\frac{ln2}{c_1}ln(A_h/A_{pl})+c_0,
\end{eqnarray}
where $c_1=\frac{\lambda^{2n_1}(1-N\lambda^2)}{1-(N\lambda^2)^{n_1+1}}$ is a positive constant and $c_0$ is a integrate constant. Although the Barrow entropy does not satisfy the standard area law, or the logarithmic correction to area law, it seams to be the well-known Boltzmann entropy $S=k_Bln\Omega$ when $A_h/A_{pl}\leftrightarrow\Omega$ and $\frac{ln2}{c_1}=1$, $c_0=0$. And the corresponding parameter $A_{pl}$ stands for the area occupied by one microscopic state, $A_h/A_{pl}$ is just the number of microscopic states of the black hole.

Since as a thermodynamic system, the thermodynamic quantities of black holes should satisfy the thermodynamical law, $dM=T_BdS_B+...$. Hence the Barrow temperature of the EPYM AdS black hole reads
\begin{eqnarray}
\frac{1}{T_B}=\frac{\partial S_B}{\partial M}=
\frac{4ln2}{c_1r_+\left(1-\Lambda r_+^2-(2q^2)^\gamma/2r_h^{4\gamma-2}\right)}.
\end{eqnarray}
When $\Lambda=0$ and $q=0$, this temperature reduces to the one of the Barrow's Schwarzschild black hole; it also can reduce to the temperature of the Barrow's RN black hole with the parameters $\Lambda=0$, $\gamma=1$. Furthermore the temperature of the Barrow's RN-AdS black hole can also be recovered with the parameter $\gamma=1$. There exists an interesting phenomena, the ratio between two entropies $\tau\equiv T_B/T=c_1\pi r_+^2/ln2$ is fully consistent with that of other Barrow's black holes. It's obviously that the Barrow's temperature and Hawking temperature both become zero in the extremal limit, $\Lambda=0$ and $r_+=2^{(\gamma-1)/(4\gamma-2)}q^{\gamma/(2\gamma-1)}$.

\begin{figure}[htp]
\includegraphics[width=0.4\textwidth]{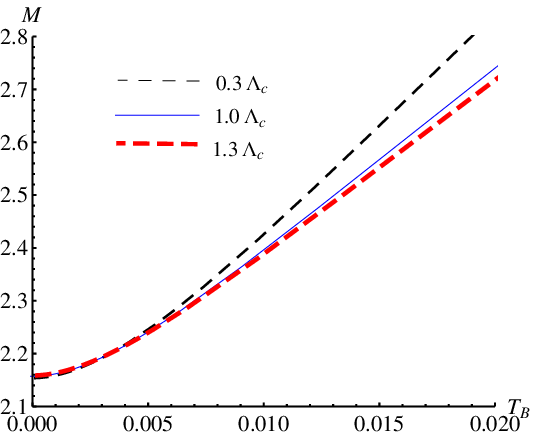}~~~~
\caption{The plots of the black hole mass vs the Barrow's temperature with the parameters $c_1=0.02,~q=0.85,~\gamma=0.8$.}\label{TB-T-M}
\end{figure}

In order to investigate the stability of the Barrow's EPYM AdS black hole, the corresponding heat capacity at the constant charge can be calculated as the following
\begin{eqnarray}
C_B=\frac{\partial M}{\partial T_B}=
\frac{2ln2\left(1-\Lambda r_+^2-\frac{\left(2 q^{2}\right)^{\gamma}}{2 r_{+}^{(4 \gamma-2)}}\right)}{c_1\left(1-3\Lambda r_+^2-\frac{(4\gamma-3)\left(2 q^{2}\right)^{\gamma}}{2 r_{+}^{4 \gamma-2}}\right)}.\label{CB}
\end{eqnarray}
The corresponding behaviours of two temperatures and the heat capacity as the functions of horizon radius are shown in Fig. \ref{T-TB-r}. For the EPYM AdS black hole with the any given value of $\Lambda$, the Barrow's temperature $T_B$ is monotonically increasing with the increasing of $r_+$, as well as the Hawking temperature $T$ with $\Lambda\geq\Lambda_c$, while as $\Lambda<\Lambda_c$ the Hawking temperature $T$ first increase to a maximum and then decreases monotonically. With the increasing of $r_+$, the heat capacity $C_B$ is positive and it also has the behaviour of the Schottky anomaly as like the standard black hole system with $\Lambda\geq\Lambda_c$. And $C_B$ will approach to zero when $T_B\rightarrow0$, in the limitation of $T_B\rightarrow\infty$ the Barrow's heat capacity $C_B$ tends to the constant $2ln2/(3c_1)$. From the viewpoint of the heat capacity, the existence of the peak indicates the Barrow's EPYM AdS black hole should have discrete energy levels microscopically. If regarding the Barrow's EPYM AdS black hole mass as the internal energy, the mass parameter should be the statistical average of these energy levels. At a enough high temperature of the Barrow's EPYM AdS black hole, the mass parameter is nearly proportional to the Barrow's temperature (see Fig. \ref{TB-T-M}). The heat capacity keeps a constant value thereafter.

\begin{figure}[htp]
\subfigure[$~~T,T_B-r_+$]{\includegraphics[width=0.4\textwidth]{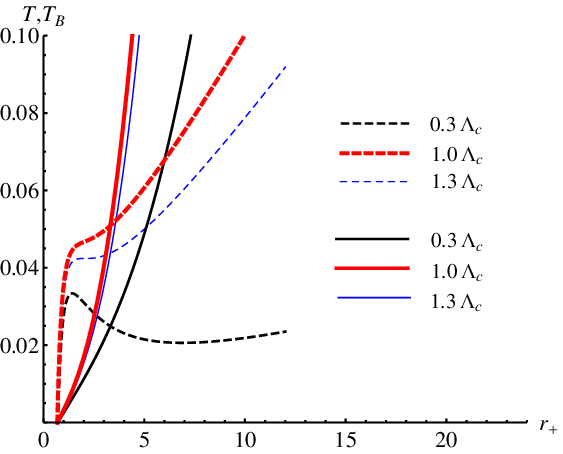}\label{T-TB-r}}~~~~
\subfigure[$~~C_B,T_B-r_+$]{\includegraphics[width=0.4\textwidth]{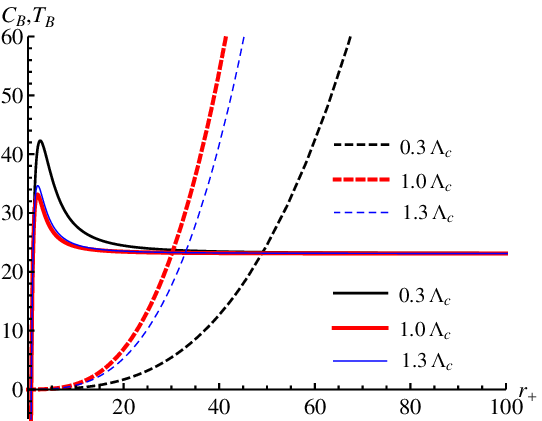}\label{CB-TB-r}}~~~~
\caption{The parameters are set to $c_1=0.02,~q=0.85$ and $\gamma=0.8$. In the left, the dashed line stands for the behaviour of $T_B-r_+$, and the solid line is the curve of $T-r_+$. In the right, the solid line stands for the behaviour of $C_B-r_+$, and the dashed line is the curve of $T_B-r_+$.}\label{T-TB-r}
\end{figure}

\section{Discussions and Conclusions}
\label{scheme4}
Based on the Barrow's idea of the fractal structure for black holes, the authors \cite{Ma2022} proposed the area of the smallest bubble sphere is the minimal increase of horizon area. The entropy is of a logarithmic form for the static spherically symmetric black holes, which can be related to the Boltzmann entropy that can reflect the microscopic structure of black holes. Based on these we found that the Barrow's EPYM AdS black hole with a any given value of $\Lambda$ is always stable and there is no phase transition as well as the standard EPYM AdS black hole with $\Lambda\geq\Lambda_c$, and as $\Lambda<\Lambda_c$ the standard EPYM AdS black hole will undergo a first-order phase transition in which a unstable intermediate-potential black hole can be survived. On the other hand, for the Barrow's EPYM AdS black hole its heat capacity exhibits a Schottky anomaly behavior, which has also been found and discussed in other gravitational system. This maybe due to the existence of discrete energy levels and restricted microscopic degrees of freedom. For low temperatures, the Barrow's heat capacity increases rapidly with the Barrow's temperature, while at high enough temperature its behaviour is like that of any usual thermodynamical systems.

\section*{Acknowledgements}

We would like to thank Prof. Ren Zhao and Meng-Sen Ma for their indispensable discussions and comments. This work was supported by the National Natural Science Foundation of China (12375050, 12075143), the Natural Science Foundation of Shanxi Province (202203021221209, 202303021211180), the Teaching Reform Project of Shanxi Datong Universtiy (XJG2022234), and the Program of State Key Laboratory of Quantum Optics and Quantum Optics Devices (KF202403).

\end{document}